# Laser-scanning optical-frequency-comb microscopy for multimodal imaging

SHIMPEI KAJIWARA,[1,2] EIJI HASE,[3] SHOTA NAKANO,[1] KEISHIRO OOTANI,[4] TOMOYA OKABE,[4] HIDENORI KORESAWA,[1,2] AKIFUMI ASAHARA,[5] KAZUMICHI YOSHII,[3,6] HIROTSUGU YAMAMOTO,[7] KAORU MINOSHIMA,[3,5] TAKESHI YASUI,[3] AND TAKEO MINAMIKAWA,[3,8,*]

[1]*Graduate School of Advanced Technology and Science, Tokushima University, 2-1 Minami-Josanjima, Tokushima, Tokushima 770-8506, Japan*
[2]*Otsuka Electronics Co. Ltd., 3-26-3, Shodai-Tajika, Hirakata-shi, Osaka, 573-1132, Japan*
[3]*Institute of Post-LED Photonics (pLED), Tokushima University, 2-1 Minami-Josanjima, Tokushima, Tokushima 770-8506, Japan*
[4]*Graduate School of Sciences and Technology for Innovation, Tokushima University, 2-1, Minami-Josanjima, Tokushima, Tokushima 770-8506, Japan*
[5]*Graduate School of Informatics and Engineering, The University of Electro-Communications, 1-5-1 Chofugaoka, Chofu, Tokyo 182-8585, Japan*
[6]*Graduate School of Advanced Science and Technology, Ryukoku University, 1-5 Yokotani, Seta Oe-cho, Otsu, Shiga 520-2194, Japan*
[7]*Center for Optical Research and Education, Utsunomiya University, 7-1-2, Yoto, Utsunomiya, Tochigi 321-8585, Japan*
[8]*Graduate School of Engineering Science, Osaka University, 1-3 Machikaneyama, Toyonaka, Osaka 560-8531, Japan*
*\*minamikawa.takeo.es@osaka-u.ac.jp*

Abstract: We introduce a novel laser-scanning optical microscopy technique that employs optical-frequency-comb (OFC) lasers. This method facilitates multimodal spectroscopic imaging by analyzing interferograms produced via a dual-comb spectroscopic approach. Such interferograms capture comprehensive light information, including amplitude, phase, polarization, frequency, and time of flight information, enabling multimodal imaging from a single measurement. We demonstrate the potential of this technique across several spectroscopic imaging applications.

## 1. Introduction

Optical microscopy is now widely used for the observation of microstructures and for the analysis of molecular and structural functions of samples, which has been applied in the fields of industry, biology, medicine, and so on[1, 2]. Furthermore, optical microscopy can employ a high numerical aperture objective lens for the tight focusing of excitation light, which can improve the spatial resolution in both plane and axial resolution by confocal configuration, molecular sensitivity by tightly confined irradiated light, imaging contrast by eliminating stray light, and also to enhance photon-molecular interactions such as nonlinear optical phenomena[3-5].

Traditional optical microscopy using tightly focused light generally observes optical intensity by employing a sample- or laser-scanning system to analyze samples via reflectance, scattering, absorbance, and laser-induced phenomena. Another contrasting method is using the optical phase, which can enhance the image contrast of highly transparent materials and thin-structured films[6, 7]. However, an interferometric measurement with tightly focused light increases measurement time due to the optical path scanning of the reference light and the wavelength scanning of monochromatic light. The long-time measurement also results in poor mechanical and thermal stability. Polarization is another important piece of information of light for analyzing birefringence and molecular orientation[8-14]. However, polarization microscopy with tightly focused light requires a mechanically driven polarization modulator, which increases measurement time. An electrically-driven polarization modulator, such as an electro-optic or photoelastic modulator, is another option to improve measurement time[15, 16]. Still, the spectral bandwidth at simultaneous measurement would be limited due to the narrow spectral response of the polarization modulators. In addition to these problems involved in conventional optical microscopy using tightly focused light, a suitable optical configuration should be individually employed to obtain each aspect of light, complicating the optical system and making measurement time-consuming.

To overcome these limitations presented in the conventional methods, in this study, we propose laser-scanning optical microscopy employing optical-frequency-comb (OFC) lasers, namely, laser-scanning OFC microscopy. The OFC laser exhibits well-stabilized amplitude, phase, and optical frequency because of the stabilization of the repetition rate and carrier-envelope offset of laser pulses[17, 18]. Using two OFC lasers with slightly different repetition rates and interfering with each other, a highly defined interferogram with an ultra-wide time span without mechanical scanning can be obtained, i.e., dual-comb spectroscopy[19-30]. The interferogram involves temporal information such as envelope, carrier, and time-of-flight information of light, enabling the characterization of samples in the time domain. The Fourier transformation decomposes the interferogram on each frequency component's amplitude and phase information, enabling the characterization of samples in the frequency domain[31-34].

Furthermore, polarization can be characterized by the respective components of light intensities and phases along each coordinate axis[35-37]. Therefore, the combination of laser scanning optical microscopy and the OFC technique enables the characterization of samples via comprehensive optical responses in the time and frequency domains with tightly focused light. In this study, we provided a proof-of-principle demonstration of laser-scanning OFC microscopy.

## 2. Principle of operation

### 2.1 Experimental setup

The experimental setup of the developed laser-scanning OFC microscopy is shown in Fig. 1. Erbium-doped fiber laser-based dual-comb system (OCLS-HS-FREP-TK, Neoark Corp.) was employed as OFC generators operating at the center wavelength of 1560 nm. One OFC generator (signal comb) with a repetition rate ($f_{rep}$) of 100 MHz was used for the illumination of samples. The other OFC generator (local comb) was used to decode the signal comb reflecting sample information via a dual-comb spectroscopic procedure, of which the repetition rate was set at a slightly different repetition rate from the signal comb ($\Delta f_{rep}$ = 688.6 Hz). These two OFC generators were well-stabilized and synchronized with each other by a feedback-control system using an electronic optical modulator installed in the local comb.

The signal comb was split into two paths: the sample path for illuminating the sample, and the reference path for generating measurement triggers and a reference signal. In the sample path, an objective lens focused the signal comb onto the sample. For reconstructing a two-dimensional spatial image of the sample, the focal spot was scanned in two dimensions using a pair of galvanometer mirrors. The signal comb, reflecting back through the same objective lens and mirrors, was then combined with the local comb using a beam splitter. A photodetector detected the overlapped signals after causing interference. Additionally, for analyzing polarization, we incorporated a polarization beam splitter after merging the signal and local combs, capturing the interference signals of the two orthogonal polarization components with two separate photodetectors. The same detection method was applied for the reference path, albeit with the path length reduced to nearly half that of the sample path, facilitating temporal distinction between the interferograms from each path.

*2.2 Analytical procedure*

Due to a slight discrepancy in the repetition rates of the signal and local combs, the temporal interference between them translated the temporal characteristics of the signal comb into a time-stretched interferogram by a factor of $f_{rep}/\Delta f_{rep}$. A Fourier transform was applied to decompose the interferogram into amplitude and phase spectra, which reflect the frequency behavior of the signal comb. By analyzing these spectra, along with the interferogram obtained through the sample path and comparing it with that from the reference path, we characterized the samples based on their temporal and frequency responses.

To assess the polarization properties of each frequency component, we determined the amplitude and phase of the orthogonal polarization components by Fourier transforming their respective interferograms. The polarization state of each frequency component is represented as a Jones vector:

$$\begin{bmatrix} E_x \\ E_y \end{bmatrix} = \begin{bmatrix} \hat{E}_x \exp(j\varphi_x) \\ \hat{E}_y \exp(j\varphi_y) \end{bmatrix}, \quad (1)$$

where $\hat{E}$ and $\varphi$ denote the amplitude and phase of the electric field components, respectively. With Eq. 1, polarization parameters such as ellipticity and azimuth angle can be analyzed, allowing the laser-scanning OFC microscopy to perform polarization analysis by directly measuring the amplitude and phase of each polarization component, thereby eliminating the need for mechanical movements.

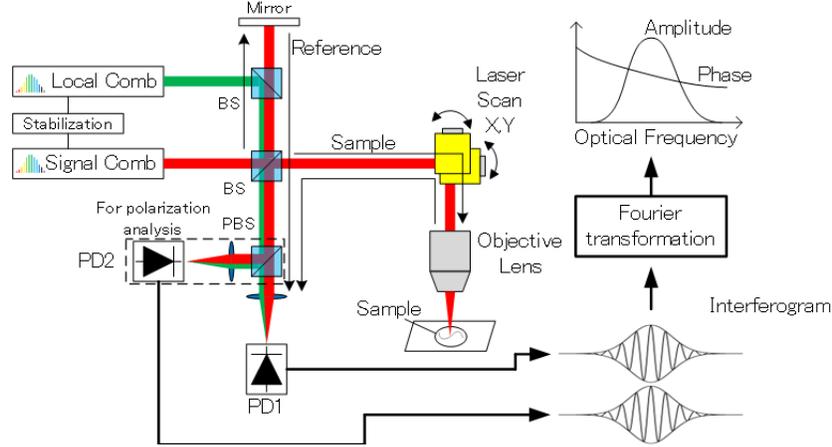

Fig. 1 Experimental setup and procedure of the laser-scanning OFC microscopy.

## 3. Results

*3.1 Fundamental characteristics of laser-scanning OFC microscopy*

### 3.1.1 Spectroscopic amplitude and phase imaging of a test chart

Firstly, we evaluated the laser-scanning OFC microscopy system using a test chart, as depicted in Fig. 2a. This chart consisted of a glass substrate coated with a chromium layer, approximately 100 nm thick. The chromium coating and the glass substrate demonstrated distinct reflective properties, providing pronounced amplitude contrast. Moreover, the thickness of the chromium layer induced a significant phase contrast. We observed the region highlighted in Fig. 2a using laser-scanning OFC microscopy, which yielded detailed interferograms for each pixel. From this data, we reconstructed spectroscopic amplitude and phase images, thereby enhancing our understanding of the imaging capabilities and performance of the developed system.

Typical interferograms from the chromium plate and the glass substrate are presented in Figs. 2b and 2c, illustrating the differences in reflectance as variations in the amplitude of the interferograms from the sample path. A closer analysis of the interferograms, shown in Figs. 2d and 2e, further reveals variations in the phase of the carriers, indicative of the phase shifts introduced by the chromium coating steps.

We applied Fast Fourier Transform (FFT) to the sample path's interferogram, with the resulting amplitude and phase spectra displayed in Figs. 2f and 2g. These spectra, centered around the light source wavelength of 1560 nm, decompose the amplitude and phase differences into their respective optical frequency components. This analysis allows for a detailed examination of the frequency-dependent reflectance and optical path differences between the glass substrate and chromium coating.

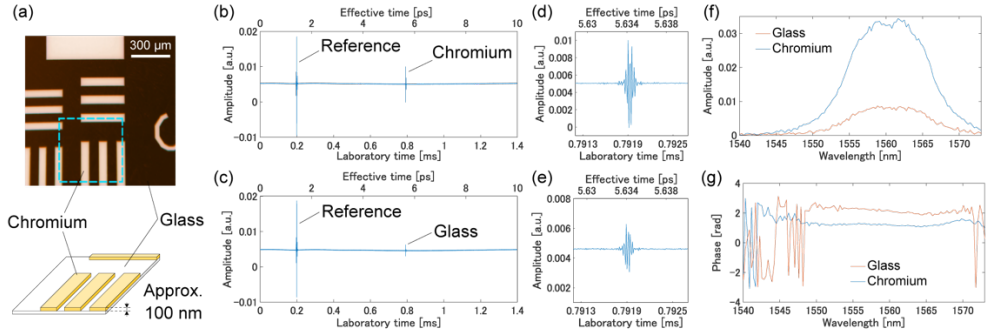

Fig. 2: Laser-scanning OFC microscopy analysis of a test chart. (a) A white light image and schematic of the test chart, with the blue dashed square indicating the imaging region depicted in Fig. 3. Interferograms displaying one complete period of the repetition rate observed at (b) a chromium film and (c) a glass substrate, with reference interferograms appearing at staggered time positions. Enlarged interferograms extracted from (d) the chromium film and (e) the glass substrate. Fourier-transformed (f) amplitude and (g) phase spectra, derived from the interferograms of both the chromium film and the glass substrate."

Typical amplitude and phase spectral images at 1560.25 nm are displayed in Fig. 3. The amplitude image highlights the contrast arising from variations in reflectance between the chromium thin film and the underlying glass substrate, as shown in Fig. 3a. During phase imaging, we observed phase variations across the plane, even when examining flat mirrors. These variations were attributed to changes in the beam path induced by laser scanning. To address this issue, we conducted phase calibration using a flat mirror prior to measuring the test chart. This calibration effectively delineated image contrasts corresponding to the step structure of chromium deposition on the glass base, as demonstrated in Fig. 3b.

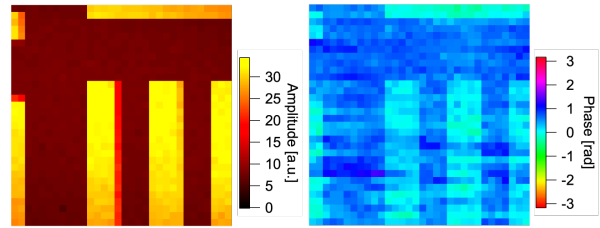

Fig. 3 Laser-scanning OFC microscopic amplitude and phase imaging of a test chart at 1560.25 nm. (a) Amplitude image showing the contrast derived from different reflectance properties. (b) Phase image displays variations due to beam path di the step structure of the chromium deposition on the glass base. The area of interest is highlighted by a blue square in Fig. 2. Image dimensions are 444 μm × 444 μm with a resolution of 32 × 32 pixels.

### 3.1.2 Phase stability

Next, we assessed the phase stability of the interferogram to evaluate the stability of the laser-scanning OFC microscope over the measurement time. Phase stability, defined as the fluctuation in the interferogram phase, depends on the robustness of the optical setup and the timing jitter between the signal comb and the local comb. We measured 100 integrated interferograms over approximately 8 minutes without laser scanning. Using FFT analysis on the interferogram, we quantified the phase's time variation at 1560.25 nm. This variation is depicted in Fig. 4. The standard deviation of the phase variation was 0.16 rad, corresponding to an accuracy of ±19.9 nm in position measurement.

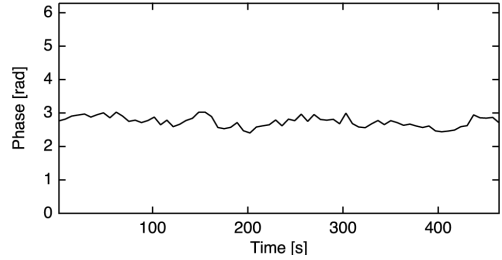

Fig. 4: Phase stability of the laser-scanning OFC microscope. The phase's time variation at 1560.25 nm was evaluated using 100 integrated interferograms over approximately 8 minutes without laser scanning. The standard deviation of the phase variation was 0.16 rad.

### 3.1.3 Spatial resolution

Spatial resolution was then evaluated using a high numerical aperture lens (NA = 0.95). The spatial resolution was calculated from the edge spread function obtained from the images of the edges of the test charts used as samples. Fig. 5a shows the bright-field image of the test chart and the area used for the spatial resolution evaluation (dashed line area). An amplitude image at 1560.25 nm observed using the laser-scanning OFC microscope is shown in Fig. 5b.

The spatial resolution was estimated from the half-width of the line spread function, which is the derivative of the edge spread function. The spatial resolution of the x-axis and y-axis were estimated to be 1.44 μm and 1.28 μm, respectively. The theoretical value is given by:

$$\delta = 0.51 \frac{\lambda}{NA}, \qquad (2)$$

where $\delta$, $\lambda$, and NA represent spatial resolution, wavelength, and numerical aperture, respectively. The theoretical spatial resolution of 0.84 μm deviates somewhat from the measured value. This discrepancy is likely due to the limited beam diameter incident on the objective pupil, which was done to increase the light source utilization efficiency, resulting in an insufficient NA. Additionally, the use of an objective lens not optimized for long wavelengths likely caused aberrations, reducing the practical spatial resolution. Despite these limitations, the spatial resolution obtained is sufficient to resolve μm-sized structures. By using an objective lens optimized for NA utilization efficiency and the wavelength used, a spatial resolution comparable to that of a conventional optical microscope can be achieved.

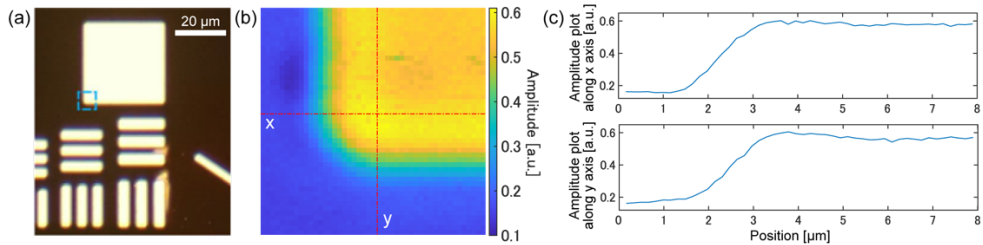

Fig. 5 Spatial resolution of the laser-scanning OFC microscope. (a) Bright-field image of a test chart. (b) Amplitude image of the area within the dashed line in Fig. 5a, obtained by the laser-scanning OFC microscope. (c) Cross-sectional amplitude plots along the x- and y-axes indicated in Fig. 5b. The image size is 7.9 μm × 7.9 μm, with a pixel resolution of 48 × 48.

## 3.2 Applications utilizing various optical information obtained by laser-scanning OFC microscopy

### 3.2.1 Phase imaging for surface profilometry

First, surface profilometry measurements of stepped specimens were performed to demonstrate the nanometer depth resolution and micrometer in-plane resolution of phase imaging. A schematic diagram of the sample is shown in Fig. 6a. Measurements with an atomic force microscope (OLS3500-PTU, Olympus) showed that $h_1$, $h_2$, and $h_3$ were 315.4 ± 12.1 nm, 122.5 ± 11.6 nm, and 50.5 ± 10.4 nm, respectively. Laser scanning OFC microscopy images are shown in Figs. 6b and 6c. As each step consisted of the same material, there was no contrast in the amplitude image but clear contrast in the phase image.

To calculate the height information from the phase information, the following equation is used:

$$h_n = \frac{\lambda}{4\pi}(\varphi_n - \varphi_{n-1}), \tag{3}$$

where $n$, $\varphi_n$, $h_n$, and $\lambda$ are the step number (1, 2, 3), the phase obtained at each step, the height of each step, and the wavelength of the light source, respectively. $\phi_0$ indicates the phase obtained at the bottom plate. The surface profile obtained from the phase image in Fig. 6c, transformed into a surface profile using Eq. 3, is shown in Fig. 6d. As a result, $h_1$, $h_2$, and $h_3$ were estimated to be 310 ± 28 nm, 126 ± 22 nm, and 50 ± 20 nm, respectively. The correlation between the surface profile measurements obtained by atomic force microscopy and laser-scanning OFC microscopy is shown in Fig. 6e. The heights measured by laser-scanning OFC microscopy agreed with those measured by atomic force microscopy.

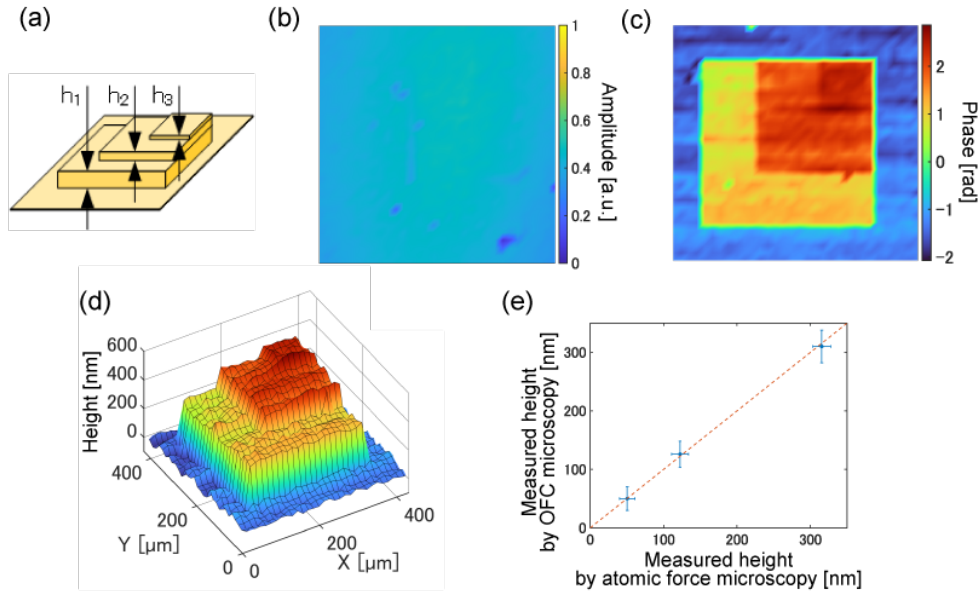

Fig. 6 Surface profilometry of the step sample by laser-scanning OFC microscopy. (a) Schematic of the step sample. (b) Amplitude image and (c) phase image obtained by the laser-scanning OFC microscope. Image dimensions are 444 μm × 444 μm with a resolution of 32 × 32 pixels. (d) Surface profilometry of the step sample transformed from the phase image. (e) Correlation of each step height between atomic force microscopy and laser-scanning OFC microscopy.

### 3.2.2 Time-of-flight imaging for surface profilometry

As mentioned in the introduction, it is difficult to find phase steps larger than the wavelength of the excitation light using the phase difference method. Laser-scanning OFC microscopy can overcome this limitation through time-of-flight measurements, which observe the time of arrival of the pulse to the detector. The difference in the time of arrival of the pulses is caused by the position of the object and the difference in the speed of light in the object, as show in Fig. 7a. This difference is reflected in the temporal position of interferograms obtained by laser-scanning OFC microscopy. By measuring the temporal position of the interferogram, it is possible to estimate the thickness and height of objects that produce phase differences longer than the wavelength.

We verified the fundamental capability of time-of-flight measurement with a laser-scanning OFC microscope using three glass samples with different heights (0.15 mm, 0.52 mm, and 0.95 mm). Typical interferograms and two-dimensional images of the glass samples are shown in Fig. 7b. In this measurement, $f_{rep}$ and $\Delta f_{rep}$ were 10 MHz and 688.4 Hz, respectively. Thus, the actual effective time of the pulse train can be calculated by multiplying the time scaling factor by the laboratory time, where the time scaling factor was 14,526 (calculated as $f_{rep}/\Delta f_{rep}$).

The time of arrival of the pulses is calculated by identifying the peak of the pulse envelope, then converted to the height of the glass $h$ using the following equation:

$$h = \frac{c}{2(n_{glass} - 1)} \Delta t \qquad (4)$$

where $c$, $n_{glass}$, and $\Delta t$ are the speed of light, the refractive index of glass, and the time of arrival of the pulse, respectively. By employing a typical relative refractive index of glass of 1.52, the glass heights were estimated to be 0.150 mm, 0.519 mm, and 0.954 mm. These results are summarized in Table 1. The estimated height measured using the laser-scanning OFC microscope matched well with those measured using calipers (0.15 mm, 0.52 mm, and 0.95 mm), as shown in Fig. 7c. In two-dimensional imaging of the glass edges by laser-scanning OFC microscopy, two-dimensional images reflecting the height of the glass were also successfully obtained, as shown in Fig. 7d.

**Table 1 Height estimation of glasses by laser-scanning OFC microscopy.**

| Glass height measured by calipers | Peak position of interferogram in laboratory time | Time of arrival of pulse in actual effective time | Estimated glass height |
|---|---|---|---|
| 0.15 mm | 0.075 μs | 0.519 ps | 0.150 mm |
| 0.52 mm | 0.262 μs | 1.802 ps | 0.519 mm |
| 0.95 mm | 0.481 μs | 3.312 ps | 0.954 mm |

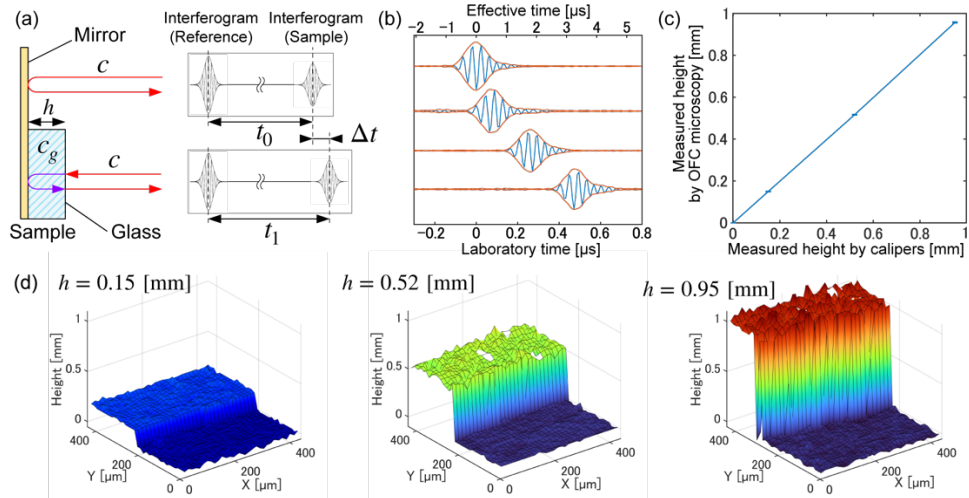

Fig. 7 Time-of-flight measurement by laser-scanning OFC microscopy. (a) Time delay in pulse arrival due to differences in the optical path length, which is also reflected in the interferogram.

(b) Typical interferograms using three glass samples with different heights (0.15 mm, 0.52 mm, and 0.95 mm). (c) Correlation of the glass heights evaluated by the laser-scanning OFC microscopy and the calipers. (d) Two-dimensional imaging of the edge of the glass samples.

### 3.2.3 Spectroscopic polarization imaging

Laser-scanning OFC microscopy can also be used for polarization measurements with an almost identical optical system. For polarization measurements, a polarizing beam splitter was installed just before the detector, and the interferograms of the orthogonal polarization components (x- and y-polarization) were acquired with separate detectors.

A proof-of-principle for polarization analysis using a laser-scanning OFC microscope was performed with a zero-order quarter-waveplate (WPQ501, Thorlabs) mounted on a mirror. Note that the amplitude and phase of the direct in-plane polarization component can be directly acquired with the laser-scanning OFC microscope. Therefore, the incident polarization can be set arbitrarily. In the present study, the ellipticity and phase difference of the incident light were set to -0.49 rad and 0.98 rad, respectively.

The amplitude ratio and phase difference between x- and y-polarization when a zero-order quarter-waveplate is used are shown in Figs. 8a and 8b. The theoretical values, which account for the incident light passing through the waveplate, reflecting off a mirror, and passing through the waveplate again, are also shown in Figs. 8a and 8b. The observed phase difference and amplitude ratio of the x- and y-polarized light agreed well with the theoretical values across both wavelength and quarter-waveplate angle.

Since the laser-scanning OFC microscope allows for the independent acquisition of amplitude and phase for each of the x- and y-polarizations, polarization parameters such as ellipticity $\chi$ and azimuth angle $\Psi$ can be calculated directly using the following equations:

$$\chi = \frac{1}{2}\sin^{-1}\left[\sin\left(2\tan^{-1}\frac{\hat{E}_y}{\hat{E}_x}\right) \cdot \sin(\varphi_y - \varphi_x)\right], \tag{6}$$

$$\Psi = \frac{1}{2}\tan^{-1}\left[\frac{2\hat{E}_x\hat{E}_y \cos(\varphi_y - \varphi_x)}{1 - \hat{E}_y^{\,2}}\right], \tag{7}$$

where $\hat{E}$ and $\varphi$ are the amplitude and phase of each polarization component. The observed polarization parameters agreed well with the theoretical values, as shown in Figs. 8c and 8d. Note that the polarization parameters are obtained directly from the phase of each polarization component, allowing the determination of which polarization component (x or y) is delayed. This makes it possible to determine whether the polarization exhibits clockwise or counterclockwise rotation from the sign of the ellipticity. In the results shown in 8c, the ellipticity exhibited a positive value, indicating that the polarization rotated in a clockwise direction.

A two-dimensional imaging result of the quarter-waveplate using the laser-scanning OFC microscope is shown in Fig. 8e. In the mirror region, which does not have the property of modulating polarization, values representing the incident polarization were obtained uniformly. In contrast, in the quarter-waveplate region, different polarization characteristics were observed compared to the mirror region. Note that in the waveplate region, uniform polarization characteristics were obtained because a waveplate with a uniform in-plane structure was used in this study. This indicates that the laser-scanning OFC microscope can analyze polarization in two-dimensional space.

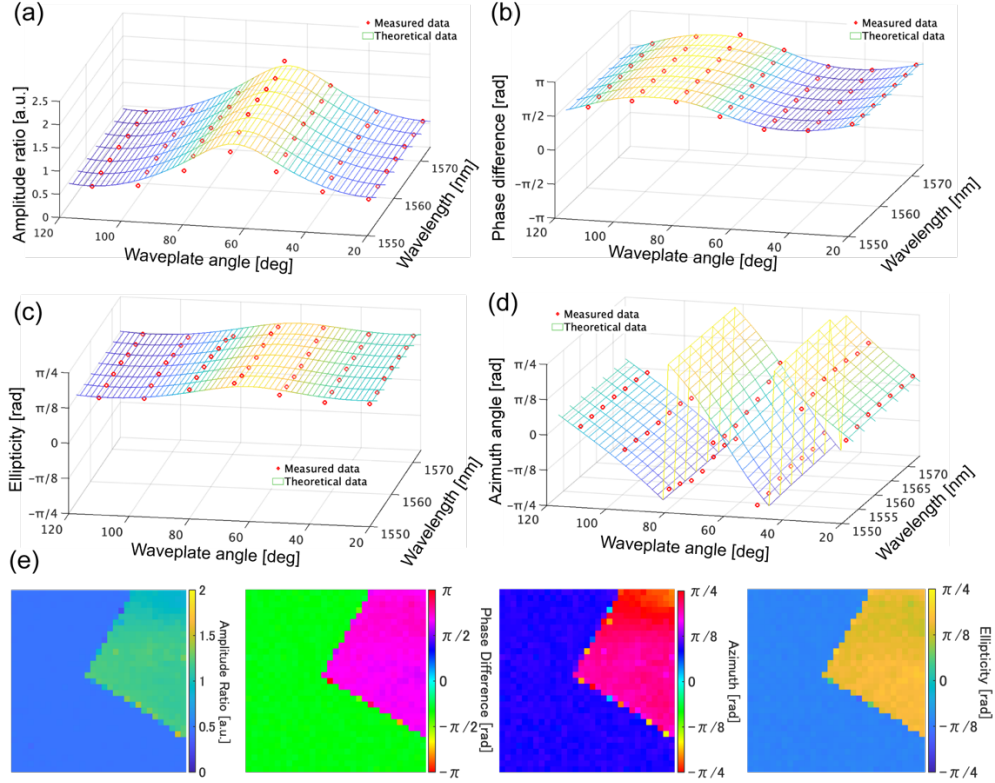

Fig. 8 Laser-scanning OFC microscopic polarization imaging of a quarter-waveplate. (a) Amplitude ratio and (b) phase difference as a function of the wavelength of incident light and waveplate angle. Estimated polarization parameters of (c) ellipticity and (d) azimuth angle as a function of the wavelength of incident light and waveplate angle. (e) Two-dimensional polarization analysis of the waveplate (upper right region) and a mirror (other region). The analyzed wavelength was 1560.25 nm.

## 4. Discussion

This study demonstrates the efficacy and versatility of laser-scanning OFC microscopy, which enables multimodal imaging using multiple aspects of light information, such as amplitude, phase, polarization, frequency, and time of flight, into a single measurement. Our findings affirm that this technology enhances the traditional capabilities of optical microscopy and opens new avenues for detailed spectroscopic imaging.

Our experimental results underscore the capability of laser-scanning OFC microscopy to handle complex multimodal imaging tasks with remarkable efficiency. The system's versatility is evidenced by its ability to analyze a broad spectrum of sample characteristics, from millimeter to nanometer-scale structures, using a single experimental setup. This ability to capture comprehensive optical parameters significantly reduces measurement times and increases throughput. Such functions are expected to fully utilize the capabilities of light, aiding in the understanding of new optical phenomena and applications in industrial metrology.

Despite its many advantages, our system also faces challenges related to its complexity and environmental sensitivity. In particular, the phase accuracy of our laser-scanning OFC microscope is ±19.9 nm, which still falls short of the sub-nanometer accuracy of established techniques such as atomic force microscopy and white light interferometry. Furthermore, the system is particularly susceptible to atmospheric disturbances due to the long optical paths required for setup on the optical bench. To improve phase stability, future optical paths should be housed or compacted to protect them from environmental disturbances. Additionally, considering the interaction between light and matter, improvements could include employing OFC lasers with shorter or longer wavelengths to verify applications in the ultraviolet, visible, and mid-infrared regions, which interact more acutely with matter.

## 5. Conclusion

We developed and demonstrated laser-scanning spectroscopic microscopy based on Fourier transform spectroscopy without the need for mechanical step-scanning in the time domain, referred to as laser-scanning OFC microscopy. Our method enables the multimodal simultaneous observation of amplitude, phase, time, and polarization with tightly focused light during two-dimensional imaging. This novel technique will serve as an effective tool for analyzing the microscopic structures and functions of materials in various scientific fields and industrial applications

**Funding**

Exploratory Research for Advanced Technology (ERATO) MINOSHIMA Intelligent Optical Synthesizer Project (JPMJER1304), Japan Science and Technology Agency (JST), Japan; Precursory Research for Embryonic Science and Technology (PRESTO) (JPMJPR17PC), JST, Japan; Grant-in-Aid for Scientific Research (B) (JP18H01901) from the Japan Society for the Promotion of Science (JSPS).

**Acknowledgments**

The authors sincerely acknowledge Dr. Takahiko Mizuno of Anritsu Corp. for his technical suggestions and discussions on OFC laser sources, and Ms. Asaka Murakami of Tokushima University for her assistance with English proofreading of the manuscript.

**Disclosures**

KS is an employee of Otsuka Electronics Co., Ltd. during this project. The other authors declare no conflicts of interest.